# Half-life of $^{31}$Si

G. D'Agostino[(1)], M. Di Luzio[(1),(2)], G. Mana[(3)] and M. Oddone[(2)]

(1) Istituto Nazionale di Ricerca Metrologica (INRIM), Unit of Radiochemistry and Spectroscopy, c/o Department of Chemistry, University of Pavia, via Taramelli 12, 27100 Pavia, Italy

(2) Department of Chemistry, University of Pavia, via Taramelli 12, 27100 Pavia, Italy

(3) Istituto Nazionale di Ricerca Metrologica (INRIM), Strada delle Cacce 91, 10135 Torino, Italy

Email of the corresponding author: g.dagostino@inrim.it

Half-life values are widely used in nuclear chemistry to model the exponential decay of the quantified radionuclides. The analysis of existing data reveals a general lack of information on the performed experiments and an almost complete absence of uncertainty budgets. This is the situation for $^{31}$Si, the radionuclide produced via neutron capture reaction recently used to quantify trace amounts of $^{30}$Si in a sample of the silicon material highly enriched in $^{28}$Si and used for the determination of the Avogadro constant. In order to improve the quality of the now recommended 157.36(26) min value, we carried out repeated observations of the $^{31}$Si decay rate via $\gamma$-ray spectrometry measurements. This paper reports the result we obtained, including details of the experiment and the evaluation of the uncertainty.

## I. INTRODUCTION

A special issue of the journal *Metrologia* was recently dedicated to the evaluation of uncertainties in radionuclide metrology [1]. The measurement of the half-life of radionuclides was one of the covered topics [2]. In particular, it was reported that for many radionuclides the scattering of the half-life values obtained in different determinations is larger than anticipated from their uncertainties. Moreover, details on the experiment and how the uncertainty was evaluated are frequently missing.

The reliability of recommended half-life values concerns different fields, including analytical techniques based on radioactivity. Since half-life correction factors are used to rescale the measured activities to a reference time, the value and the uncertainty propagation of the adopted half-life might significantly bias or affect the uncertainty of the quantified analyte.

This was the case of a recent measurement of the $^{30}$Si mole fraction, $x(^{30}$Si$)$, of a silicon single crystal highly enriched in $^{28}$Si used for the determination of the Avogadro constant [3]. The relative contribution to the combined uncertainty of $x(^{30}$Si$)$ of the adopted $t_{1/2}$ value of the $^{31}$Si [4] is 35 %. Although the data used to derive the recommended value are not discrepant, a close check of the referenced papers confirms the lack of details on experiments and uncertainty evaluation.

Since an improvement of the present knowledge of the $^{31}$Si half-life might be useful, we carried an experiment based on neutron activation and $\gamma$-counting. Details of the measurement are reported in the following, including the measurement model, the corrections and the uncertainty budget of the result.

## II. MODEL

The way we propose to measure the $^{31}$Si half-life, $t_{1/2}$, is to perform repeated observations of its exponential decay rate via $\gamma$-ray spectrometry, the single observation being a sequence of successive counts of the 1266.1 keV $\gamma$-photons emitted during the radioactive decay of a $^{31}$Si source [2].

In details, the 1266.1 keV $\gamma$-peak count rate at the beginning of the $i^{th}$ count of the $j^{th}$ sequence, $C_{ij}(t_{d\,ij})$, starting at $t_{d\,ij}$ and lasting $t_{c\,ij}$, is

$$C_{ij}(t_{d\,ij}) = \frac{\lambda\, n_{ij}}{(1-e^{-\lambda t_{c\,ij}})} \delta_{ij} f_{ij}, \qquad (1)$$

where $\lambda = \ln(2)/t_{1/2}$ is the $^{31}$Si decay constant, $n_{ij}$, $\delta_{ij}$ and $f_{ij}$ are the net count of the 1266.1 keV $\gamma$-peak, the dead time correction and the pile-up correction of the $i^{th}$ count of the $j^{th}$ sequence, respectively.

The $n_{ij}$ value is obtained by subtracting the background count, $b_{ij}$, from the gross count, $g_{ij}$, of the $\gamma$-peak. The $\delta_{ij}$ and $f_{ij}$ values depend on the total input count rate, $C_{T\,ij}$, and correct the lost counts due to dead time and pile-up, respectively. In the case of $\lambda\, t_{c\,ij} \ll 1$ and $t_{dead\,ij} \ll t_{c\,ij}$,

$$\delta_{ij} = \frac{t_{c\,ij}}{t_{c\,ij} - t_{dead\,ij}} \quad \text{and} \quad f_{ij} = e^{\mu(t_{dead\,ij}/t_{c\,ij})}, \qquad (2)$$

where $t_{dead\,ij}$ is the dead time and $\mu$ is a (positive) constant, hereafter called pile-up factor, inversely proportional to the resolution time of the electronic system. Usually, $t_{dead\,ij}/t_{c\,ij}$ is defined as the relative dead time.

The $C_{ij}(t_{d\,ij})$ value is proportional to the $^{31}$Si activity, with the constant of proportionality being the detection efficiency of the $\gamma$-peak, $\eta$, which takes into account (i) the source-detector geometry, (ii) the intrinsic efficiency for the particular gamma-ray energy, (iii) the incomplete absorptions due to Compton scattering and (iv) the probability of emission of the observed gamma radiation.

If the $j^{th}$ sequence of counts is performed without moving the source and under stability conditions of the detection system, $\eta_j$ is constant and the variation of the count rate, (1), with time can be modeled by

$$C_{ij}(t_{d\,ij}) = \hat{C}_j(t_{d1j})\, e^{-\lambda t_{d\,ij}} + \varepsilon_{ij}, \qquad (3)$$

where $\hat{C}_j(t_{d1j})$ is the expected value of the count rate at the starting time of the first count of the $j^{th}$ sequence, $t_{d\,1j}$, and $\varepsilon_{ij}$ is the error term.

### III. EXPERIMENT

The detection system used in this study consists of a coaxial Ge detector ORTEC® GEM50P4-83 (relative efficiency 50 %, resolution 1.90 keV FWHM at 1332 keV) connected to a digital signal processor ORTEC DSPEC 502. The end cap of the detector is placed inside a low-background graded lead shield located in an underground laboratory with a temperature-controlled atmosphere of 23 °C.

Preliminary experiments were carried out to test the stability of the detection system and to measure the pile-up factor.

The stability was checked by recording a sequence of 750 successive counts of the 661.8 keV $\gamma$-photons emitted during the radioactive decay of a $^{137}$Cs source. The uninterrupted observation time was 375 h.

The pile-up correction factor was measured by performing several counts of the 661.8 keV $\gamma$-photons of the $^{137}$Cs source with the addition of a $^{152}$Eu source located at different positions with respect to the detector end cap to change the total input count rate, $C_T$ [5]. Two sequences of 34 counts were repeated with the $^{152}$Eu source in six different positions while two sequences of 34 counts were repeated without the $^{152}$Eu source. The overall observation time was 320 h and consisted of 476 counts.

In both cases, the $^{137}$Cs source was fixed at about 60 mm from the detector end cap to keep constant the detection efficiency of the 661.8 keV $\gamma$-peak. In addition, the counting window, $t_c$, was adjusted on line to reach a 0.1 % relative uncertainty of the 661.8 keV net peak due to counting statistics.

The measurement of the $^{31}$Si half-life was carried out with a 6 g sample of a high purity silicon single crystal. This sample was used as a $^{31}$Si source produced by the activation of $^{30}$Si via the neutron capture reaction $^{30}$Si(n,$\gamma$)$^{31}$Si. Twelve repeated neutron activations and observations of the decay rate were performed. On average, a single sequence lasted 400 min and consisted of 26 successive counts. The overall observation time was 169 h and consisted of 310 counts.

Each neutron irradiation lasted 3 h and was carried out at the central thimble of a 250 kW TRIGA Mark II reactor. After activation, the $^{31}$Si source was fixed at about 10 mm from the end cap of the detector. The $^{31}$Si decay rate was observed by recording a sequence of counts of the 1266.1 keV $\gamma$-photons collected with a constant detection efficiency of the $\gamma$-peak. The counting window was adjusted on line to achieve a 1 % relative uncertainty of the 1266.1 keV net peak due to counting statistics.

## IV. DATA ANALYSIS AND DISCUSSION

The net counts of the $\gamma$-peaks were estimated from the collected spectra using the WAN32 analysis algorithm of the ORTEC® Gamma Vision software [6]. The results in the (few) cases of $\gamma$-peak located in a region automatically deconvoluted by the software were rejected to maintain the best accuracy.

### A. Stability and pile-up of the detection system

The stability and pile-up factor of the detection system was checked and measured, respectively, by processing the $\gamma$-spectra recorded during the sequences of counts of the $^{137}$Cs source.

The 661.8 keV $\gamma$-peak count rate of the source measured at the $i^{th}$ count of the $j^{th}$ sequence, $C_{ij}(t_{d\,ij})$, and corrected for decay at the starting time of the first count of the first sequence, $t_{d\,11}$, is

$$C_{ij}(t_{d11}) = \frac{C_{ij}(t_{d\,ij})}{e^{-\lambda(t_{d\,ij}-t_{d11})}}, \tag{4}$$

where $C_{ij}(t_{d\,ij})$ is obtained from (1) using the $^{137}$Cs decay constant, $\lambda$ ($t_{1/2} = 30.08$ y), and the net count of the 661.8 keV $\gamma$-peak, $n_{ij}$. Since the $\gamma$-photons counting was performed with a constant detection efficiency, the expected value of $C_{ij}(t_{d\,11})$, $\hat{C}(t_{d\,11})$, is constant as well.

In addition, the following model applies:

$$C_{\text{pu}\,ij}(t_{d\,11}) = \hat{C}(t_{d\,11})\,e^{-\mu(t_{\text{dead}\,ij}/t_{c\,ij})} + \varepsilon_{ij}, \tag{5}$$

where $C_{\text{pu}\,ij}(t_{d\,11}) = \dfrac{\lambda\,n_{ij}}{(1-e^{-\lambda t_{c\,ij}})\,e^{-\lambda(t_{d\,ij}-t_{d\,11})}}\,\delta_{ij}$ is the $\gamma$-peak count rate measured at the $i^{\text{th}}$ count of the $j^{\text{th}}$ sequence uncorrected for pile-up and corrected for decay at $t_{d\,11}$.

The single sequence of 750 counts collected for the stability test was performed at a fixed relative dead time, 0.077. Therefore, $j = 1$ and $i = 1, 2, \ldots, 750$; the $\delta_{i1}$ and $f_{i1}$ values are constant. In the case of a stable detection system, the ratio of the $C_{i1}(t_{d\,11})$ value obtained from (4) to the average of the $C_{i1}(t_{d\,11})$ values, $C_{\text{ave1}}(t_{d\,11})$, must scatter due to counting statistics around the expected unity value without a deterministic drift. This was the case of the ratios obtained in this study and plotted in Figure 1.

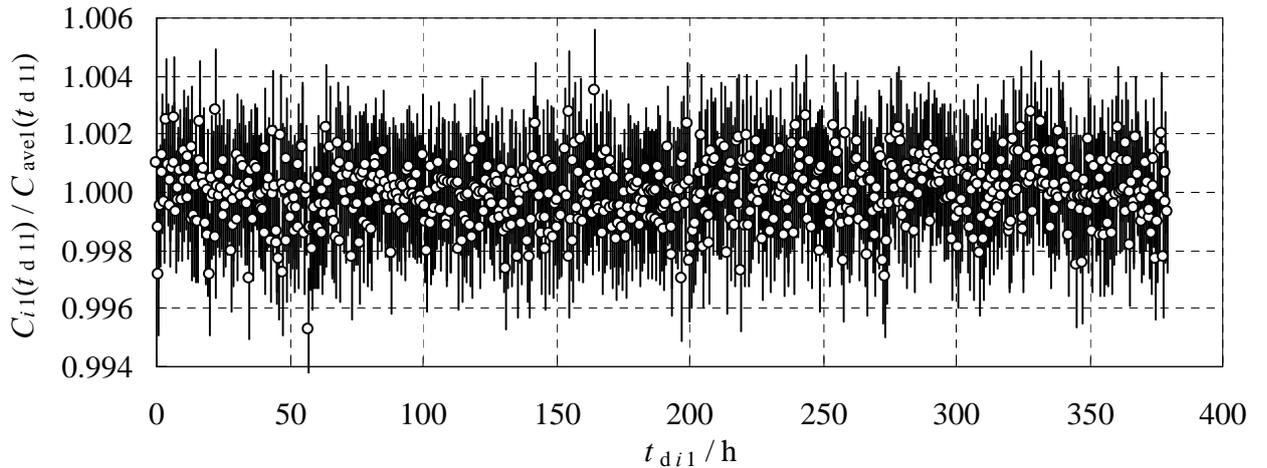

FIG. 1. The normalized measured count rate of the $^{137}$Cs source corrected for decay as a function of time. The error bars indicate a 95 % confidence interval due to counting statistics.

The average relative uncertainty due to counting statistics of the 661.8 keV $\gamma$-peak net count and evaluated by the WAN32 analysis algorithm is $1.048 \times 10^{-3}$ whereas the experimental standard deviation of the ratios is $1.064 \times 10^{-3}$. This close agreement proves that the detection system doesn't introduce effects which alter the variability of the data due to their Poisson nature.

The fourteen sequences of 34 counts collected for the measurement of the pile-up factor were performed with the detection system working at seven different relative dead times, i.e. 0.077 (1$^{\text{st}}$ and 8$^{\text{th}}$ sequence), 0.116 (2$^{\text{nd}}$ and 9$^{\text{th}}$ sequence), 0.153 (3$^{\text{rd}}$ and 10$^{\text{th}}$ sequence), 0.189 (4$^{\text{th}}$ and 11$^{\text{th}}$ sequence), 0.220 (5$^{\text{th}}$ and 12$^{\text{th}}$ sequence), 0.253 (6$^{\text{th}}$ and 13$^{\text{th}}$ sequence) and 0.288 (7$^{\text{th}}$ and 14$^{\text{th}}$ sequence). Therefore, $j = 1, 2, \ldots, 14$ and $i = 1, 2, \ldots, 34$; the $\delta_{ij}$ and $f_{ij}$ values are changing according to (2).

Linear least squares regression is applied to estimate the pile-up factor, $\mu$, by fitting a straight line to the natural logarithm of (5) averaged with respect to the 34 counts of the $j^{th}$ sequence, $C_{pu\,avej}(t_{d11})$, and normalized to the $C_{pu11}(t_{d11})$ value. Expressly, the measured variable, $y_{ij}$, is $\ln[C_{pu\,avej}(t_{d11})/C_{pu11}(t_{d11})]$, the independent variable $x_{ij}$ is $t_{dead\,avej}/t_{c\,avej}$ and the estimated parameters $\beta_1$ and $\beta_2$ are $\ln[\hat{C}(t_{d11})/C_{pu11}(t_{d11})]$ and $-\mu$. Since the main contribution to the uncertainty of $C_{pu\,ij}(t_{d11})$ is the 0.001 relative uncertainty of the 661.8 keV net peak, the standard deviation of the error term of $y_{ij}$, $\varepsilon_{ij}/C_{pu\,avej}(t_{d11})$, is constant and close to $0.001/\sqrt{34}$.

The straight line fitting the data and the residuals are shown in the upper and lower graph of Figure 2, respectively. The estimated pile-up factor, $\mu$, is $4.45(8) \times 10^{-3}$. Here and hereafter the uncertainties ($k = 1$) in parentheses apply to the last respective digits.

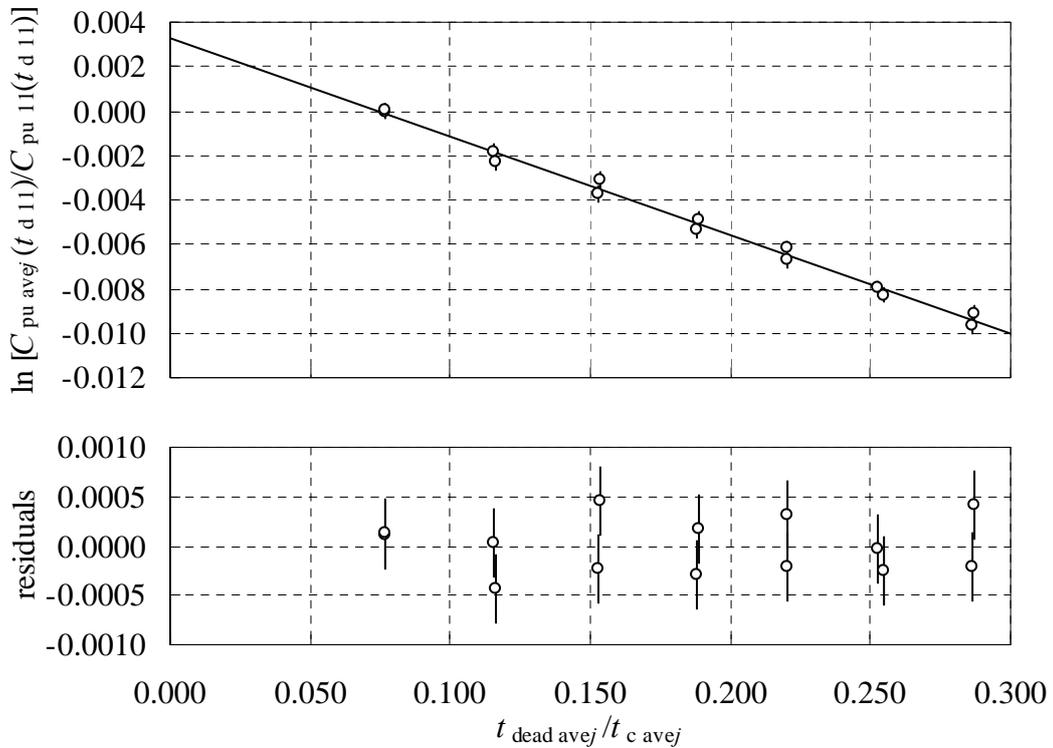

FIG. 2. The straight line (upper graph) and the residuals (lower graph) obtained by fitting the data collected to determine the pile-up factor. The error bars indicate a 95 % confidence interval due to counting statistics.

### B. The decay constant of $^{31}$Si

The decay constant of $^{31}$Si was determined by processing the $\gamma$-spectra recorded during the twelve sequences of counts of the neutron activated $^{31}$Si source. The number of performed counts in a sequence, $N$, depended on the $^{31}$Si activity at the start of the observation. Specifically, the $N$ values were 26 (1st sequence), 29 (2nd sequence), 25 (3rd sequence), 28 (4th sequence), 26 (5th sequence), 32 (6th sequence), 25 (7th sequence), 14 (8th sequence), 27 (9th sequence), 31 (10th sequence), 25 (11th sequence) and 22 (12th sequence).

The relative dead time varied, on average, from 0.30 ($t_c = 460$ s) to 0.06 ($t_c = 33$ min) whereas the maximum and the minimum relative, 0.31 and 0.03, were reached at the first and last count of the

6th sequence which lasted 9.2 h. Furthermore, the ratio of the background count, $b$, to the net count, $n$, of the $^{31}$Si 1266.1 keV $\gamma$-peak, $\alpha$, was almost constant and in the range between 0.010 and 0.035.

The 1266.1 keV $\gamma$-peak count rate at the beginning of the $i^{th}$ count of the $j^{th}$ sequence, $C_{ij}(t_{d\,ij})$, is computed according to (1), where $j = 1, 2, \ldots, 12$ and $i = 1, 2, \ldots, N$. The $\delta_{ij}$ and $f_{ij}$ values are obtained from (2).

Linear least squares regression is applied to estimate $\lambda$, and the corresponding $t_{1/2}$, by fitting twelve straight lines to the twelve series of data obtained by taking the natural logarithm of (3) normalized to the $C_{1j}(t_{d1j})$ value. Specifically, the measured variable, $y_{ij}$, is $\ln[C_{ij}(t_{d\,ij})/C_{1j}(t_{d1j})]$, the independent variable $x_{ij}$ is $t_{d\,ij}$ and the estimated parameters $\beta_{1j}$ and $\beta_2$ are the twelve intercepts, $\ln[\hat{C}_j(t_{d1j})/C_{1j}(t_{d1j})]$, and the shared slope, $-\lambda$, respectively. Since $y_{ij}$ depends on $\lambda$, the solution is obtained iteratively until convergence.

The main contribution to the uncertainty of $C_{ij}(t_{d\,ij})$ is the 0.01 relative uncertainty of the 1266.1 keV net peak. Therefore, the standard deviation of the error term of $y_{ij}$, $\varepsilon_{ij} / C_{ij}(t_{d\,ij})$, is constant and equal to 0.01 as well.

The regression analysis was performed with the Levenberg Marquardt algorithm of the OriginLab® software using the line model in global multi-data fit mode [7]. The twelve straight lines fitting the data and the residuals plotted with respect to $t_{d\,ij} - t_{d\,1j}$ are shown in the upper and lower graph of Figure 3, respectively. The resulting $^{31}$Si decay constant, $\lambda$, is $4.4103 \times 10^{-3}$ min$^{-1}$ with an associated (fitting) uncertainty of $0.0046 \times 10^{-3}$ min$^{-1}$.

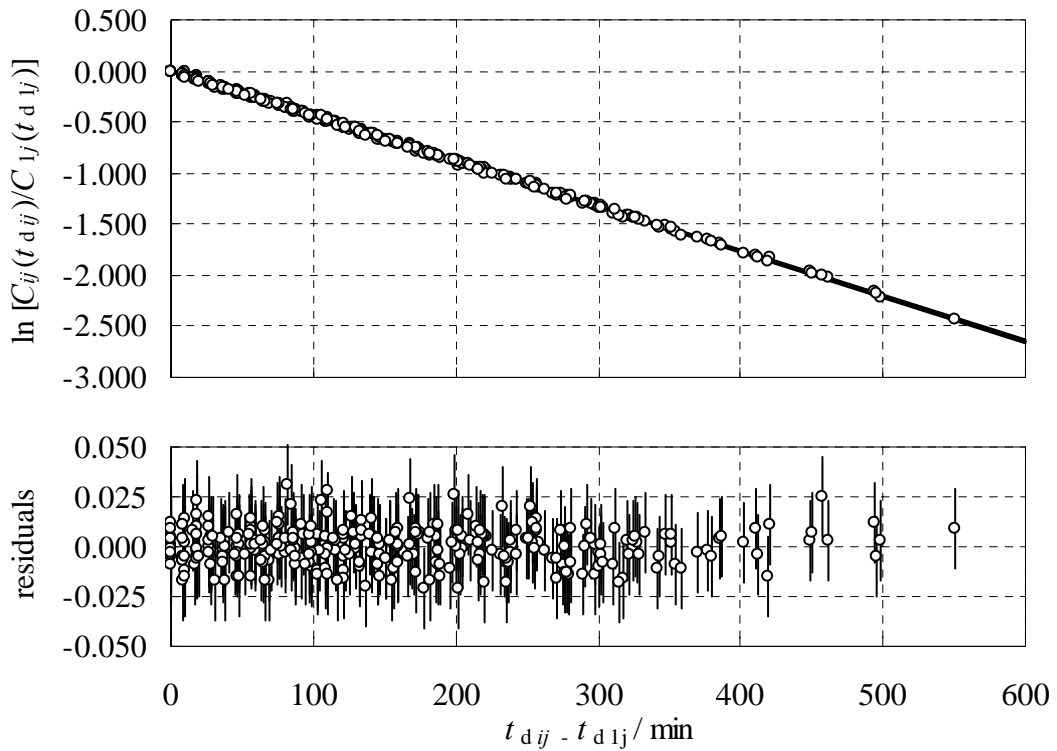

FIG. 3. The twelve straight lines (upper graph) and the residuals (lower graph) obtained by fitting the data collected to determine the $^{31}$Si decay constant. The error bars indicate a 95 % confidence interval due to counting statistics.

## C. Uncertainty

In the case of a perfect linear relationship between the count rate and the activity of the source, the residuals are expected to be pure stochastic and due to the Poisson nature of counting. Thus, the only contribution to the uncertainty of $\lambda$ would be the fitting. However, medium- (cyclic) and low-frequency (drift) deviations must carefully be considered because they can significantly affect the uncertainty also when they are hidden in the residuals. The most important sources of these deviations are (i) instability of the detection efficiency, (ii) pile-up and dead time, (iii) geometrical changes in source positioning, (iv) spectral interference and (v) background signals [2].

The effect of counting statistics can be anticipated using equations (34) and (37) reported in [2] and based on the number, $N$, of activity measurements, $A$, and associated relative uncertainties, $\sigma(A)/A$, repeated at regular time during an observation lasting $T$. Although in this study the activity (count rate) measurements, were not equally spaced in time, the resulting relative uncertainty, $\sigma(\lambda)/\lambda$, with $T = 400$ min, $N = 26$ and $\sigma(A)/A = 0.01$ is about 0.003 for a single sequence. The corresponding value for the mean of twelve of these sequences, 0.0009, is close to the evaluated (fitting) relative uncertainty, 0.0010.

The results of the stability test didn't show any cyclic variations or drift of the detection efficiency system added to the expected experimental 0.001 relative standard deviation. Thus, 0.0003 is a reasonable upper limit for their contribution to $\sigma(A)/A$.

The pile-up correction, $f$, in the case of a relative dead time 0.30 and 0.06 is 1.0013 and 1.0003, respectively. A 30 % error of the (corrected) 0.001 relative variation of $f$ between the start and end of a sequence leads to an average relative uncertainty, $<\sigma(A)/A>$, i.e. at $t = 0$ and $t = 400$ min, of 0.00015. Since the counting and dead times have a negligible uncertainty, the contribution of the dead time correction, $\delta$, to $\sigma(A)/A$ is negligible as well.

During a single observation, the position of the $^{31}$Si source was fixed respect to the detector, therefore the effect of geometrical changes in source positioning is absent.

Given the high-purity of the silicon sample used as a source, the 1266.1 keV $\gamma$-peak was free from spectral interference. The Compton continuum underlying the $\gamma$-peak was approximately flat and due to natural background. To achieve the 0.01 relative uncertainty due to counting statistics, the net count of the $\gamma$-peak, $n$, reached a value close to $1 \times 10^3$ in every count. Hence, the effect of a possible and undisclosed departure of the $\gamma$-peak from the Gaussian shape is considered constant during the observation. In addition, the upper limit for the relative background contamination of the fitted $\gamma$-peak area is fixed to 10 % of the maximum $\alpha$ value, i.e. 0.0035. A 30 % variation of the relative background contamination between the start and end of a sequence leads to an average relative uncertainty, $<\sigma(A)/A>$, i.e. at $t = 0$ and $t = 400$ min, of about 0.00053.

In agreement with [2], a propagation factor $2/\lambda T$ is assumed for the $\sigma(A)/A$ due to (i) variations or drift of the detection efficiency system, (ii) pile-up correction and (iii) background contamination. The corresponding relative uncertainty, $\sigma(\lambda)/\lambda$, is 0.00034, 0.00018 and 0.00060.

In summary, an overall relative uncertainty of 0.0013 is evaluated for $\lambda$. The main contributors are counting statistics, 68 %, and background contamination, 22 %.

## D. Comparison with literature data

The presently adopted $^{31}$Si half-life value, 157.36(26) min [4], is the weighted average of ten selected values reported in high quality peer-reviewed journals: 163.0(58) min [8], 157.5(5) min [9], 157.2(6) min [10], 158.4(12) min [11], 159(1) min [12], 157.3(5) min [13], 157.1(7) min [14], 155.5(10) min [15], 157.3(13) min [16] and 170(10) min [17]. These values are shown in Figure 4 and compared with the result obtained in this study, i.e. 157.16(20) min.

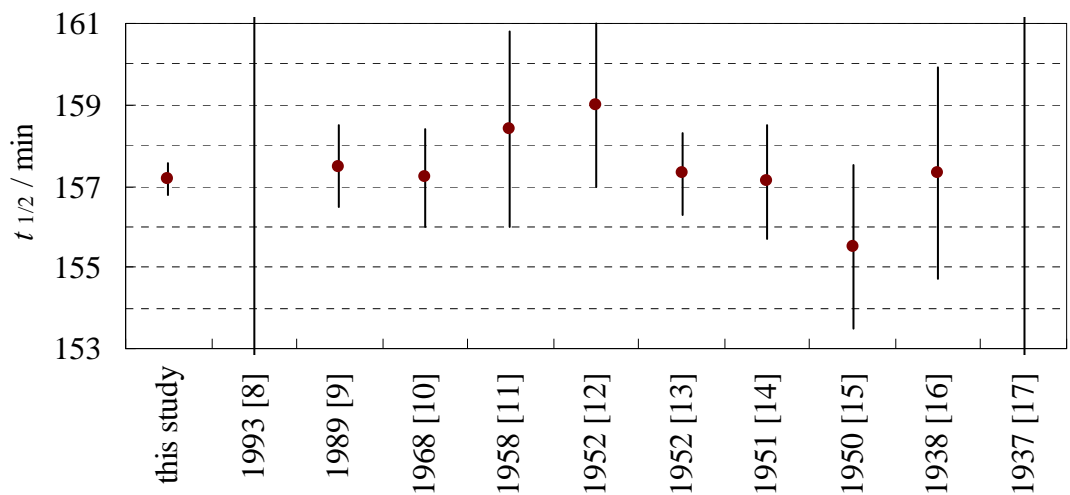

FIG. 4. The literature $^{31}$Si half-life values compared with the result obtained in this study. The reference and the publication year are reported. The error bars indicate a 95 % confidence interval.

## V. CONCLUSIONS

The $^{31}$Si half-life was determined with repeated measurements of the activity of a $^{31}$Si source activated using neutron irradiation. The value was obtained by applying linear least squares regression of data collected in different decay observations. The possible effects that might affect the result such as instability of the detection system, pile-up, dead-time and background have been limited and, whenever possible, corrected for. The relevant uncertainties were evaluated and propagated following the suggestions reported in [2].

The resulting $^{31}$Si half-life value is in agreement with the presently recommended value, which is derived from data acquired between the 40's to the 90's and reported in literature with poor information concerning details of experiments and complete absence of uncertainty budget.

## ACKNOWLEDGMENTS

This work was jointly funded by the Italian ministry of education, university, and research (awarded project P6-2013, implementation of the new SI) and by the European Metrology Research Programme (EMRP) participating countries within the European Association of National Metrology Institutes (EURAMET) and the European Union.